# Absence of magnetism in the superconductor $Ba_2Ti_2Fe_2As_4O$: Insights from inelastic neutron scattering measurements and ab initio calculations of phonon spectra

*Mohamed Zbiri[a,\*], Wentao Jin[b,#], Yinguo Xiao[c], Yunlei Sun[d], Yixi Su[b], Sultan Demirdis[b], and Guanghan Cao[d]*

[a]Institut Laue-Langevin, 71 avenue des Martyrs, Grenoble Cedex 9, 38042, France

[b]Jülich Centre for Neutron Science JCNS at Heinz Maier-Leibnitz Zentrum (MLZ), Forschungszentrum Jülich GmbH, Lichtenbergstraβe 1, D-85747 Garching, Germany

[c]Jülich Centre for Neutron Science JCNS and Peter Grünberg Institut PGI, JARA-FIT, Forschungszentrum Jülich GmbH, D-52425 Jülich, Germany

[d]Department of Physics, Zhejiang University, Hangzhou 310027, China

**ABSTRACT**

$Ba_2Ti_2Fe_2As_4O$ is a self-doped superconductor exhibiting a $T_c \sim 21.5$ K and containing, distinctively with respect to other Fe-based superconductors, not only [$Fe_2As_2$] layers but also conducting [$Ti_2O$] sheets. This compound exhibits a transition at $T^* \sim 125$ K which has tentatively been assigned in the literature to a possible density-wave order. However, the nature of this density wave (whether it is a charge- or spin-induced) is still under debate. Magnetism in $Ba_2Ti_2Fe_2As_4O$ has never been experimentally confirmed, which raises the question whether this superconductor might be non-magnetic or exhibiting a very weak magnetism. Here, we report evidence from inelastic neutron scattering (INS) measurements and ab initio calculations of phonon spectra pointing towards a possible absence of magnetism in $Ba_2Ti_2Fe_2As_4O$. The INS measurements did not reveal any noticeable magnetic effects in $Ba_2Ti_2Fe_2As_4O$, within the accessible *(Q, E)* space. The effect of magnetism on describing phonon spectra was further investigated by performing ab initio calculations. In this context, non-magnetic calculations reproduced well the measured phonon spectra. Therefore, our results indicate a non-magnetic character of the superconductor $Ba_2Ti_2Fe_2As_4O$.

# I. INTRODUCTION

The discovery of high-temperature superconductivity (HTSC) in the Fe-pnictide materials[1,2] paved the road for a tremendous progress towards the understanding of the phenomenon of superconductivity and its interplay with magnetism,[3-11] and phonon dynamics,[12-23] in these systems. The emergence of HTSC in the Fe-pnictides can either be induced chemically by a targeted doping of a parent compound,[24-26] or mechanically by applying external pressure.[27,28] However, recently, HTSC was discovered in a new iron-based oxypnictide superconductor $Ba_2Ti_2Fe_2As_4O$ ("22241"), exhibiting a bulk SC at Tc (superconducting transition temperature) ~ 21.5 K, and interestingly, subjected to an effective self-doping property.[29-32] This offers an alternative and efficient route for inducing superconductivity, instead of achieving it mechanically or chemically via element substitution. The self-doping stems from an inter-layer electronic interaction since this compound contains not only the same $[Fe_2As_2]$ layers, as in other Fe-based superconductors, but additionally another conducting $[Ti_2O]$ sheets, which makes it very distinctive. For this material, a transition at T* = 125 K has been identified from the electrical resistivity and magnetic susceptibility measurements[25]. It was tentatively ascribed by Raman scattering and optical spectroscopy studies to a possible density-wave (DW) transition in the Ti sublattice.[33,34] Further, Mössbauer measurements did not evidence the occurrence of any long-range magnetic ordering below the T* originating from the Fe sublattice in $Ba_2Ti_2Fe_2As_4O$.[35]

Previous studies on different oxygen-free Fe-based superconductors, as well as oxypnictides, highlighted the possibility that the Cooper pairing could be mediated by exchange of antiferromagnetic spin fluctuations.[7-11] However, the presently studied superconducting compound, $Ba_2Ti_2Fe_2As_4O$, seems not to be subject to such magnetic effect. Inelastic neutron scattering (INS) offers a unique tool to probe phonon dynamics over the full Brillouin zone, without any selection rule, which helps exploring phonons and their possible coupling and/or interplay with magnetic degrees of freedom. In this context we have previously studied different Fe-based pnictides using INS to collect phonon spectra, and in a synergistic way our neutron data were systematically accompanied by ab initio lattice dynamical calculations for the sake of the analysis and interpretation.[36-38] Our previous works allowed us to contribute in building-up a spin-phonon picture of the previously discovered Fe-based superconductors, by establishing the occurrence of spin-lattice coupling. In this paper, we aim at pursuing our work on phonon dynamics in Fe-pnictides by combining INS and ab initio calculations to measure and simulate, respectively, phonon spectra of $Ba_2Ti_2Fe_2As_4O$. To the best of our knowledge, only the zone-center phonons of $Ba_2Ti_2Fe_2As_4O$ have been studied via Raman spectroscopy.[33]

The aim of this paper is twofold: (i) perform INS measurements to collect phonon spectra – or generalized density of states[39,40] - of $Ba_2Ti_2Fe_2As_4O$ over an extended temperature range (2-300 K) in order to explore any signature of DW could be observed, and, (ii) carry out ab initio lattice dynamical calculations to reproduce phonon spectra and therefore gain some insights into the effect of neglecting or considering spin-polarization on describing phonon spectra in $Ba_2Ti_2Fe_2As_4O$.

This paper is organized as follows: the experimental and computational details are provided in Section-II and Section-III, respectively. Section-IV is dedicated to the presentation and discussion of the results, and conclusions are drawn in Section-V.

**II. EXPERIMENTAL DETAILS**

Polycrystalline sample of $Ba_2Ti_2Fe_2As_4O$ was synthesized by solid-state reaction method, as described in Ref. 29, and displays similar macroscopic properties as the annealed sample as reported there. The quality of the powder sample was checked by x-ray diffraction (XRD) on a Huber x-ray diffractometer with Cu $K_\alpha$ radiation at room temperature. The Rietveld refinement of the diffraction pattern at 300 K (Fig. 1(a)) reveals that a minor phase of $BaFe_2As_2$ ("122", 4.9%) coexists with the "22241" main phase (95.1%) in the sample. In addition, the XRD pattern of the sample was also collected at low temperatures using a cryostat. Even at 13 K, which is the lowest temperature value we could reach, the "22241" main phase still adopts the same space group, $I_{4/mmm}$, as at the room temperature. The XRD pattern at 13 K (Fig. 1(b)) can be well refined using the same structure model of the main phase. The temperature dependence of the lattice constants, *a* (b) and *c*, of the "22241" phase are plotted in Fig.1 (right panel). Both *a* and *c* show a clear kink around the superconducting transition temperature, $T_c$. In addition, for the in-plane lattice constant, *a*, a tiny hump is observed around *T\**. This might be due to slight structural modulation associated with the density-wave transition in the Ti sublattice, as revealed by Raman scattering and optical spectroscopy studies[29,32].

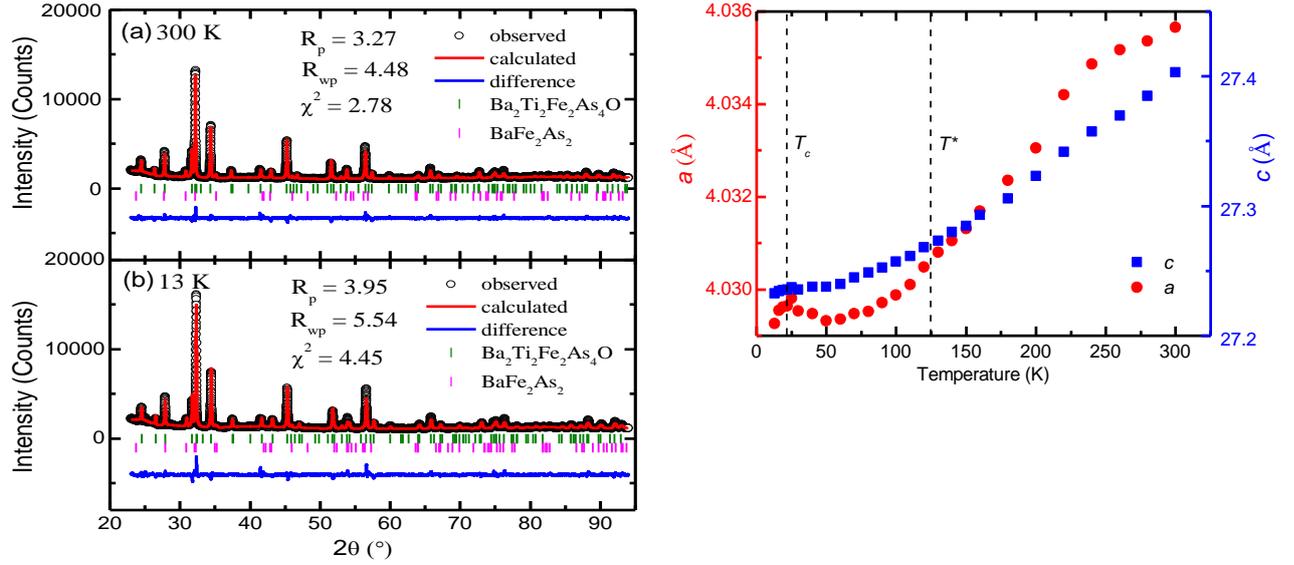

*Figure 1* (Color online) Left: the XRD refinement pattern for the polycrystalline sample at 300 K (a) and 13 (K) (b). The vertical bars at the bottom indicate the Bragg reflection positions from the $Ba_2Ti_2Fe_2As_4O$ main phase and the $BaFe_2As_2$ minor phase, respectively. The lowest curve is the difference between the observed and the calculated patterns. Right: the temperature dependence of the lattice constants, a (b) and c, of the "22241" phase. The dashed vertical lines mark the superconducting transition ($T_c$) and the density-wave transition ($T^*$).

The temperature-dependent INS measurements were performed at the Institut Laue-Langevin (ILL) (Grenoble, France) on ~ 1 gram powder sample of $Ba_2Ti_2Fe_2As_4O$. We used the thermal neutron time-of-flight spectrometer IN4C to collect phonon spectra in the temperature range 2 - 300K, in the down-scattering regime (neutron energy-loss mode), using different incident neutron wavelengths $\lambda_i$=1.11 Å ($E_i$=66.4 meV), 1.45 Å ($E_i$ = 38.9 meV), or 2.22 Å ($E_i$ = 16.6 meV). The data treatment and analysis was done using standard ILL procedures and software tools. Standard corrections including detector efficiency calibration and background subtraction were performed. Contribution from an empty aluminium container was subtracted. A standard vanadium sample was used to calibrate the detectors.

The *Q*-averaged, one-phonon generalized phonon density of states (GDOS) was obtained using the incoherent approximation in the same way as in previous works dealing with phonon dynamics in Fe-pnictides.[36-38] In the incoherent one-phonon approximation, the measured scattering function *S(Q,E)*, as observed in inelastic neutron experiments, is related to the phonon generalized density of states $g^{(n)}(E)$, as seen by neutrons, as follows:[41,42]

$$g^{(n)}(E) = A < \frac{e^{2W_i(Q)}}{Q^2} \frac{E}{n_T(E) + \frac{1}{2} \pm \frac{1}{2}} S(Q,E) > \qquad (1)$$

With:

$$g^{(n)}(E) = B \sum_i \left\{ \frac{4\pi b_i^2}{m_i} \right\} x_i g_i(E) \qquad (2)$$

where the + or − signs correspond to energy loss or gain of the neutrons, respectively, and $n_T(E)$ is the Bose-Einstein distribution. *A* and *B* are normalization constants and $b_i$, $m_i$, $x_i$, and $g_i(E)$ are, respectively, the neutron scattering length, mass, atomic fraction, and partial density of states of the i$^{th}$ atom in the unit cell. The quantity between < > represents suitable average over all *Q* values at a given energy. $2W(Q)$ is the Debye-Waller factor. The weighting factors for various atoms in the units of barns/amu are[43]: O: 0.2645, Ti: 0.0908, Fe: 0.2081, As: 0.0734 and Ba: 0.0246.

In addition, polarized neutron diffraction measurements were performed at the diffuse scattering cold-neutron spectrometer DNS[44] at the Heinz Maier-Leibnitz Zentrum (Garching, Germany), with an incident wavelength of 4.544 Å. The polycrystalline sample was put inside a thin aluminum sample container which was fixed to the cold tip of the sample stick of a $^4$He close cycle cryostat. Data were collected for 49 hours at 30 K and 16 hours at 150 K, respectively. The polarization analysis was performed via the *XYZ* method[45], by which the nuclear coherent, spin incoherent and magnetic scattering cross sections can be separated.

## III. Computational Details

The starting $Ba_2Ti_2Fe_2As_4O$ structure used in the calculations was the experimentally refined one under the tetragonal phase (a = b = 4.0276 Å and c = 27.3441 Å); with 2 formula units (22 atoms) per unit cell.[29] The refined structure (space group I4/mmm [$D^{17}_{4h}$]), contains 6 crystallographically inequivalent atoms (1 Ba, 2 As, 1 Fe, 1 Ti and 1 O). In order to determine appropriately all force constants, the supercell approach was used for the lattice dynamics calculations. Thus, a 3×3×1 supercell (SC) was constructed ($a^{sc} = b^{sc}$ = 12.0828 Å and $c^{sc}$ = 27.3441 Å), containing 18 formula units (198 atoms).

Calculations were performed using the projector-augmented wave formalism[46,47] of the Kohn-Sham density functional theory[48,49], within the generalized gradient approximation (GGA), implemented in the Vienna *ab-initio* simulation package (VASP).[50] The GGA was formulated by the Perdew–Burke–Ernzerhof (PBE) density functional.[51] The valence electronic configurations of Ba, As, Fe, Ti and O as used for pseudo potential generation are $5s^25p^66s^2$, $4s^24p^3$, $3d^74s^1$, $3d^34s^1$, and $2s^22p^4$, respectively. The Gaussian broadening technique was adopted and all results were well converged with respect to *k*-mesh and energy cutoff for the plane wave expansion. A plane wave energy cutoff of 600 eV was used, and the integrations over the Brillouin zone were sampled on a 6×6×2 grid of k-points generated by Monkhorst-pack method[52], for the supercell phonon calculations. The break condition for the self-consistent field (SCF) and ionic loops were set to $10^{-8}$ eV and $10^{-5}$ eV.Å$^{-1}$, respectively. Total energies and Hellmann-Feynman forces were calculated for 26 structures resulting from individual displacements of the symmetry inequivalent atoms in the supercell, along the six inequivalent Cartesian directions (±$x$, ±$y$ and ±$z$). Phonon density of states[39,40] and Raman frequencies were extracted from subsequent calculations using the direct method[53] as implemented in the PHONON software.[54]

## IV. Results and Discussion

We performed temperature dependent INS measurements at 2, 30, 80, 160, and 300 K. The Bose-factor-corrected *S(Q,E)* plots for Ba$_2$Ti$_2$Fe$_2$As$_4$O at 2 and 300 K, using different neutron incident energies, are shown in Figure 2. The use of a higher incident energy allowed to explore a wider *(Q,E)* space in terms of energy transfer up to 56 meV (E$_i$ = 66 meV) and Q-range, $Q_{max}$ ~ 10 Å$^{-1}$. Whereas, a smaller incident energy (E$_i$ = 17 meV) helped to explore a lower *Q*-range, down to $Q_{min}$ ~ 0.6 Å$^{-1}$, relevant to magnetism, with a better energy resolution up to an energy transfer of 14 meV. Our measurements do not show any detectable magnetic signature in the attainable *(Q,E)* range of IN4C. There is an increase of the inelastic intensity as a function of temperature and *Q,* consistent with a phonon-like behavior.

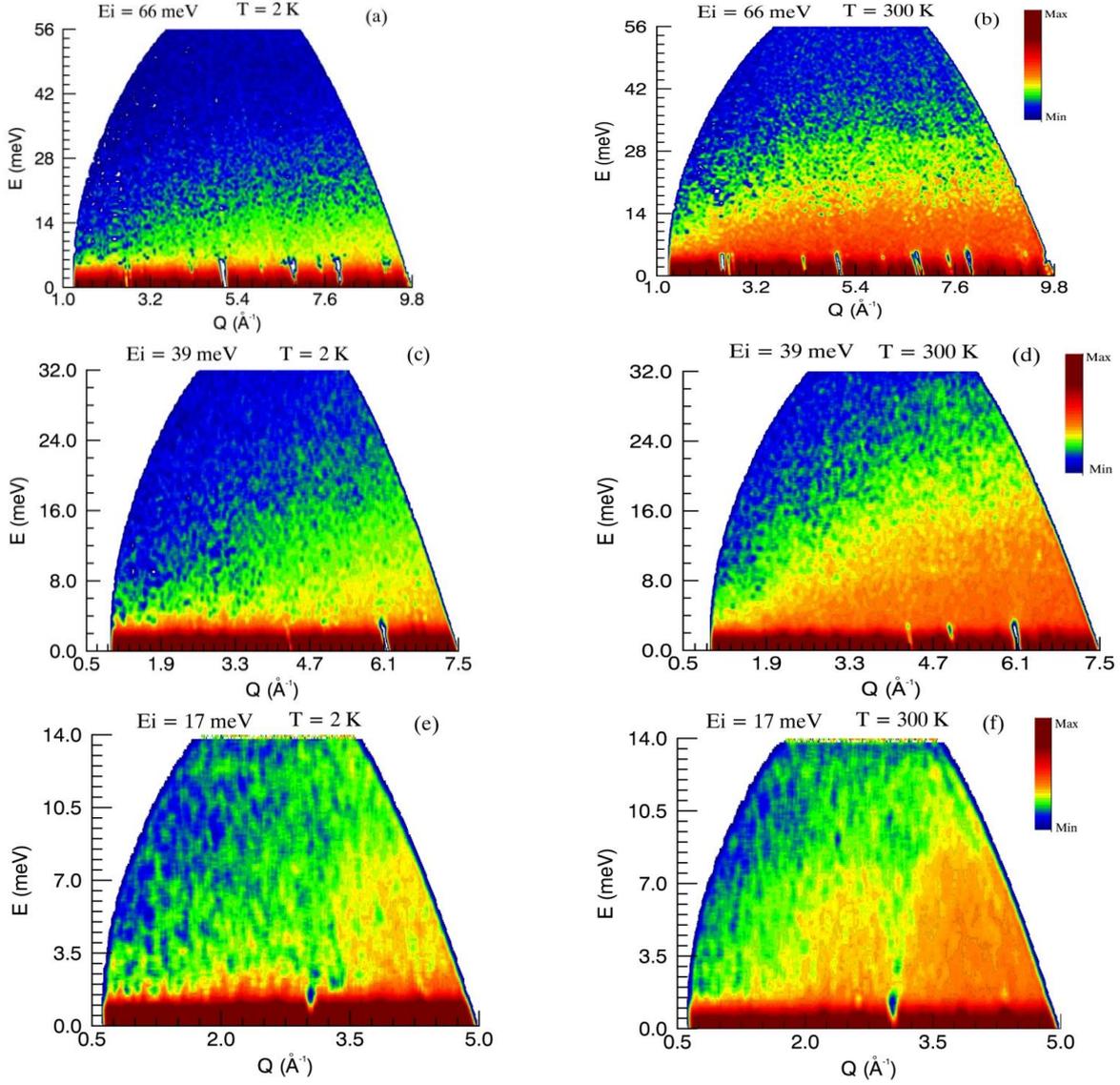

**Figure 2** *(Color online) The dynamic structure factor S(Q,E) for $Ba_2Ti_2Fe_2As_4O$ at 2 K ((a), (c) and (e)), and 300K ((b), (d), and (f)). Three incident energies were used: (a) and (b) $E_i$=66 meV ($\lambda_i$=1.11 Å), (c) and (d) $E_i$=39 meV ($\lambda_i$=1.45 Å), and, (e) and (f) $E_i$=17 meV ($\lambda_i$=2.22 Å).*

Phonon spectra, in terms of the generalized density of states[30,31] can be extracted using the S(Q,E) function (Cf. Section II). Figure 3 depicts the temperature dependence of the phonon spectra of $Ba_2Ti_2Fe_2As_4O$ at 2, 30, 80, 160, and 300 K. Our measurements show that the temperature variation has only a little effect on the evolution of the phonon spectra of $Ba_2Ti_2Fe_2As_4O$. Although this observation is quite similar to earlier studies on other Fe-based superconducting compounds (Ref[36-38] and references therein), no conclusion could definitely be made on a possible unconventional behavior of $Ba_2Ti_2Fe_2As_4O$, dedicated theoretical treatments are clearly required – which is beyond the scope of the present work.

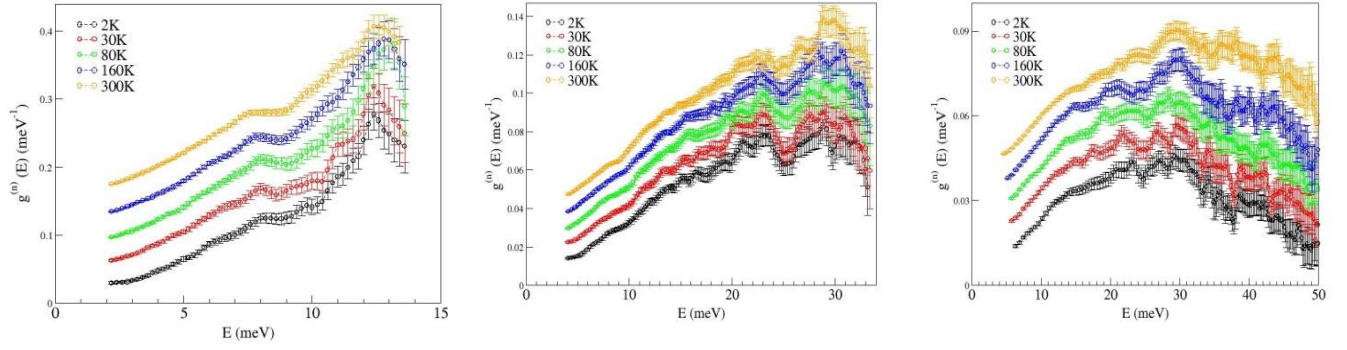

*Figure 3* (Color online) The temperature dependence of the phonon spectra GDOS of $Ba_2Ti_2Fe_2As_4O$ using different incidence neutron wavelengths: 2.22 Å ($E_i$=17 meV) (left), 1.45 Å ($E_i$=39 meV) (middle) and 1.11 Å ($E_i$=66 meV) (right). Spectra are vertically shifted to help distinguishing the observed features.

The maximum attainable energy transfer on IN4C, offering an energy resolution reasonably enough to have a global view on the spanning of phonon modes, amounts to ~ 50 meV, using an incident neutron energy $E_i$=66 meV ($\lambda_i$=1.11 Å). This did not allow capturing the vibrational feature due to Ti-O stretching modes, which are expected to occur at higher frequencies, beyond the experimental energy window. A better energy resolution is ensured when using longer neutron incident wavelengths, at the price of reducing further the attainable energy transfer range (the measurements are performed in the neutron energy-loss mode). Phonon spectra collected using incident neutron energies $E_i$=39 meV ($\lambda_i$=1.45 Å) and $E_i$=17 meV ($\lambda_i$=2.22 Å) provide a better view on distinguishable phonon features, especially those lying in the low-frequency range. In order to have a full view on the phonon spectrum and to fulfill the aims abovementioned as with respect to magnetism in $Ba_2Ti_2Fe_2As_4O$, we performed DFT-based lattice dynamical calculations to accompany our INS measurements. Figure 4 compares our ab initio calculated phonon spectra with the INS data at 2 and 300K. As the measurements were performed using different incident neutron energies (wavelengths) to ensure a balance between phase space coverage and energy resolution, the calculated phonon spectrum was convolved with a Gaussian function (Fig. 4) with a specific width in order to mimic the instrumental resolution for each incident wavelength setting. The good agreement between the calculated phonon spectra and the measurements can be appreciated, in peak positions, intensities and shapes, specifically in the case of incident energy of 17 meV ($\lambda_i$=2.22 Å). Data collected using the shortest wavelength (1.11 Å) offers an extended energy range up to 50 meV, which is an instrumental limit in the used neutron energy loss setting, and the agreement with simulated spectra is not as good as in the case of $\lambda_i$=2.22. Therefore, the validity and adequacy of the model calculation can be supported by considering the excellent agreement with the phonon spectra collected using a neutron incident wavelength of 2.22 Å ($E_i = 17$ meV), as well as with that taken at 1.45 Å ($E_i = 39$ meV).

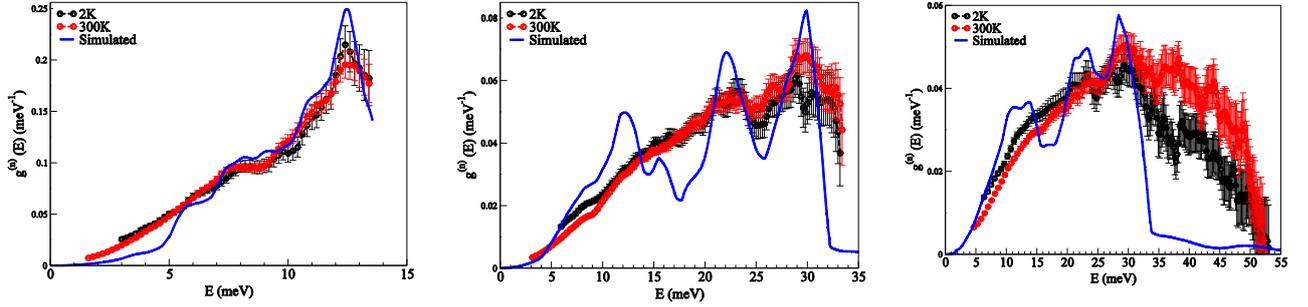

*Figure 4 (Color online) Comparison between ab initio calculated (solid line) and measured phonon spectra (dashed lines with symbols) of $Ba_2Ti_2Fe_2As_4O$ at 2 and 300 K, using different incidence neutron wavelengths: 2.22 Å (left), 1.45 Å (middle) and 1.11 Å (right). The simulated spectra were convolved with a Gaussian of full width at half maximum of 3.5, 2.5 and 1.5 meV in order to describe the effect of energy resolution in the IN4C measurements using wavelengths 1.1, 1.45 and 2.22 Å, respectively.*

The calculated phonon spectrum of $Ba_2Ti_2Fe_2As_4O$, as well as its partial atomistic, components are shown in Figure 5. As abovementioned, the vibrational feature around 72 meV could not be captured in the present INS neutron energy-loss measurements. A closer look at the partial contributions allowed to find that this feature is mainly due to O atoms, and to a much lesser extent to Ti, leading to Ti-O stretching modes (mass difference of Ti and O makes it that O exhibits a much stronger amplitude of motion). With Ba atom as the heaviest one in $Ba_2Ti_2Fe_2As_4O$, its vibrational signature is mostly located at the low-frequency range around 8 and 10 meV. $Ba_2Ti_2Fe_2As_4O$ includes two symmetry inequivalent sites occupied by As atoms. Their partial phonon densities of states reflect not only this nonequivalence, but also their bonding character in the crystal. Indeed, one Arsenide (As1) forms the [$Fe_2As_2$] layer, and this can be seen through the overlap of the partial densities of states of As1 and Fe which span basically a frequency range up to 35 meV. Whereas As2 exhibits a more localized density of states, mostly around 20 meV and seemingly decoupled from those of As1 and Fe. The localized behavior of As2 and Ba partial phonon densities of states points towards the ionic-like behavior of As2 and Ba, which is not implausible given that these two atom types are located between the layers and sheets formed by [$FeAs_1$] and [$Ti_2O$], respectively.

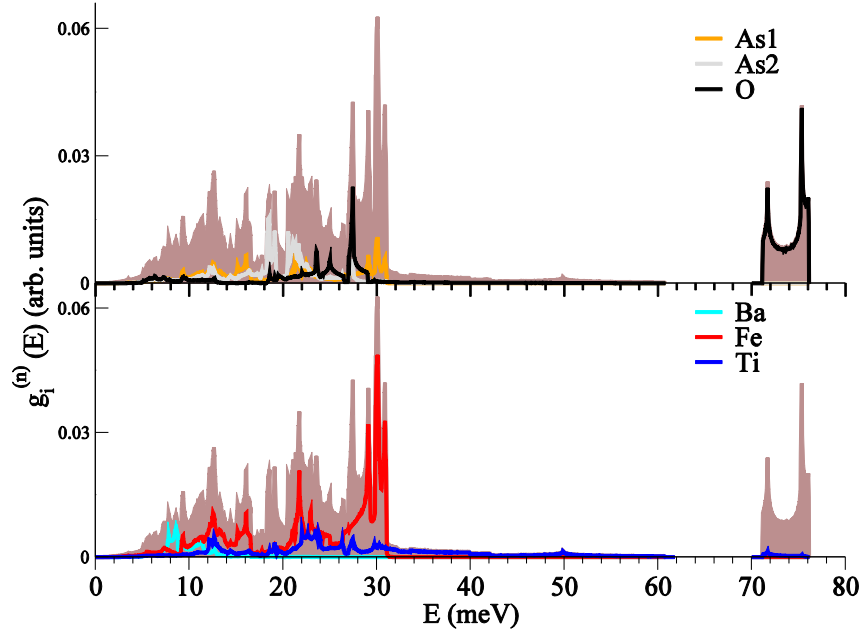

*Figure 5 (Color online) The calculated ab initio neutron-weighted partial density of states, $g_i^{(n)}$ (i=Ba, As, Fe, Ti, O), for the various atoms in $Ba_2Ti_2Fe_2As_4O$. As1 and As2 denote the two symmetry inequivalent sites occupied by As in the structure. The brown shaded area represents the total density of states in order to appreciate both position and intensity contribution of the individual atomistic components.*

Raman modes can also be extracted from our lattice dynamical calculations. Table 1 compares our calculated Raman frequencies of $Ba_2Ti_2Fe_2As_4O$ from the presently used DFT-based direct method approach,[53,54] and available experimental Raman data from Ref. 33 as well as other calculations based on linear response theory from the same work[33]. We notice the excellent agreement between the two types of lattice dynamical calculations, and the observed values, which further validates our model calculation.

| Sym | Eg(R) | A1g(R) | Eg(R) | Eg(R) | A1g(R) | A1g(R) | B1g(R) | Eg(R) |
|---|---|---|---|---|---|---|---|---|
| Exp[a] | 61,6 | 79,9 | 114,1 | 147,2 | 160,6 | 194,5 | 195,6 | - |
| Calc[b] | 62,1 | 79,7 | 135,9 | 146,5 | 184,6 | 199,4 | 200,7 | 228,4 |
| Calc[c] | 62,5 | 80,9 | 129,3 | 148,8 | 171,6 | 196,1 | 201,6 | 236,5 |

**Table 1** Comparison of calculated and measured Raman shifts ($cm^{-1}$) of $Ba_2Ti_2Fe_2As_4O$:
[a] Experimental values taken from reference Referenc[33].
[b] Calculated values based on linear response theory, taken from reference[33].
[c] Present work: calculated values based on the direct method approach (cf. Section III).

The simulated phonon spectra shown in Figure 4 and Figure 5, as well as our derived Raman frequencies gathered in Table 1, were extracted from non-magnetic lattice dynamical calculations performed on the experimentally refined structure. We have also performed other calculations[55,56] where possible effects of spin-polarization and/or electron correlation[32], via an on-site Hubbard-based correction[57,58], were considered. It comes out that only the presently reported phonon spectrum and Raman frequencies agree well with the measurements, reflecting therefore their correctness and the validity of the underlying model calculation. Indeed, including the spin polarization degree-of-freedom in the calculations based on the experimental structure[29], with or without taking into account electron correlation via DFT+U procedure,[57,58] resulted in a phonon spectrum showing imaginary frequencies, signature of structural instabilities, or in the best case, a phonon spectrum and Raman frequencies not matching our INS measurements and available Raman data[33]. We also found that calculated phonon spectra and Raman frequencies using optimized geometries deviate from the observations, reflecting the sensitivity of phonon dynamics to subtle changes in the crystal structure.[56]

|  | a, b (Å) | c (Å) | V (Å$^3$) |
|---|---|---|---|
| **Exp** | 4.0276 | 27.3441 | 443.564 |
| **NM or SP** | 4.0360 | 26.9884 | 439.622 |
| **SP$^{Hub}$** | 4.1860 | 26.6714 | 467.352 |
| **NM$^{Hub}$** | 4.0758 | 26.7161 | 443.812 |

***Table 2*** *Comparison of relaxed and experimentally refined[29] lattice parameters of $Ba_2Ti_2Fe_2As_4O$. NM and SP denote non-magnetic and spin-polarized geometry optimization calculations, respectively. In both cases, a Hubbard correction[57] (Hub) was considered or neglected. Without applying a Hubbard correction, the spin-polarized case (SP) relaxes to a non-magnetic structural solution (NM).*

Table 2 and Table 3 gather structural parameters and estimated magnetic moments, respectively, from the different model calculations mentioned above. When neglecting the effect of electron correlation, the spin-polarization (SP) case relaxes to a non-magnetic (NM) structural solution (Table 2), with a zero magnetic moment on both Ti and Fe sites ("Opt" in Table 3). Expectedly, including explicitly on-site electron correlation[57] resulted in the emergence of magnetic moments on the Fe and Ti sites in both the optimized structure ("Opt$^{Hub}$") and the experimental structure ("Exp$^{Hub}$"). A similar case is found from calculations performed on the experimental structure with a noticeable difference in terms of a non-zero, but weak, magnetic moments on the Ti and Fe sites without considering electron correlation effect ("Exp"). We notice that the calculations provide two slightly different magnetic moments on the Ti site, which is of one crystallographic type.

|  | **Exp** | **Exp$^{Hub}$** | **Opt** | **Opt$^{Hub}$** |
|---|---|---|---|---|
| **Ti** | 0.066, 0.035 | 0.8, 0.937 | 0.0 | 1.069, 1.239 |
| **Fe** | 0.584 | 2.815 | 0.0 | 3.02 |

*Table 3 Estimated magnetic moments ($\mu_B$) per Ti and Fe site from our spin-polarized DFT-based calculations performed on both the experimentally-refined structure (Exp)[29] and on the relaxed one (Opt), with (Hub) or without a Hubbard correction[57]. We notice the zero value in the case of relaxed geometry without Hubbard correction since, in this case, structure relaxes to a non-magnetic ground state geometry (Table 2).*

The fact that only non-magnetic calculations, performed on the experimentally refined structure[29] reproducing the best our INS-based phonon spectra and Raman data[33] points towards a non-magnetic nature, or a very weak and complex magnetic interactions (if any) in $Ba_2Ti_2Fe_2As_4O$. Considering spin-polarization and/or on-site electron correlation seems to destabilize both the crystal and electronic structures[59]. Comparing to the experimentally refined structural parameters[29], the spin-polarized case including the on-site electron correlation (Table 2) resulted in overestimating the crystal volume by ~ 5%, and an in-plane (*ab*) expansion of ~ 4%, while the *c*-axis is reduced by ~ 2.5%. This might explain the structural instabilities reflected in the imaginary frequencies we mentioned above. From the analysis of the other model calculations we adopted, it seems that a correct description of the axial separation governing the inter-layer interaction is very crucial in describing a coupled picture of both electronic and structural structures in $Ba_2Ti_2Fe_2As_4O$. Therefore, we speculate that this material might exhibit a charge density wave (CDW) rather than a spin density wave (SDW)-like character below $T^* = 125$ K. This by no means is contradictory with the fact that non-negligible magnetic moments are found on Fe and Ti sites when on-site electron correlation correction is included in the calculations. Indeed, imposing an on-site Coulomb interaction reduces the metal (Fe, Ti) – ligand (O, As) hybridization and leads to a localized picture which could "artificially" raise magnetic moments on Fe and Ti site. This wouldn't fully correspond to a metallic behavior of $Ba_2Ti_2Fe_2As_4O$[34,59].

In addition to our non-polarized neutron INS measurements at IN4C, we performed polarized neutron diffraction measurements on DNS to attempt getting a clearer insight into the possible static magnetism in $Ba_2Ti_2Fe_2As_4O$. Figure 6 shows the difference of the nuclear (a) and magnetic (b) scattering components separated by xyz-polarization analysis between 30 K (well below $T^*$) and 150 K (above $T^*$). The several dip-peak features in Fig. 6(a) arise from the shifting of the strong nuclear reflections while cooling. The net magnetic intensity at 30 K, after subtracting the background at 150 K, as shown in Fig. 6(b), is almost constant and close to zero, confirming the absence or extreme weakness of the magnetism blow $T^*$, if any, in $Ba_2Ti_2Fe_2As_4O$. This is well consistent with the conclusion drawn above that non-magnetic calculations can better reproduce the measured phonon spectra.

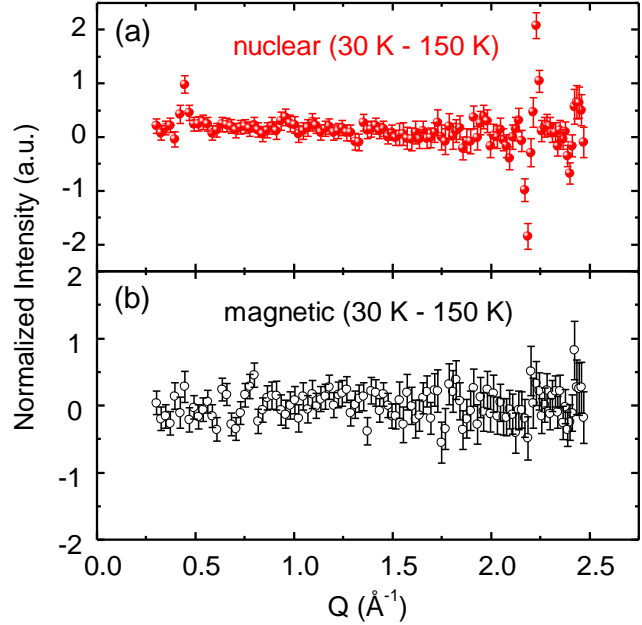

**Figure 6** *(Color online) The difference of the nuclear (a) and magnetic (b) scattering component of $Ba_2Ti_2Fe_2As_4O$ separated by xyz-polarization analysis between 30 K and 150 K.*

To summarize, our results point towards a non-magnetic character of the self-doping superconductor $Ba_2Ti_2Fe_2As_4O$, favoring a CDW instead of a SDW scenario as the origin of the transition at $T^*$. It is worth noting that no indications of a symmetry lowering expected for a SDW order were observed below $T^*$ in $Ba_2Ti_2Fe_2As_4O$ by Raman spectroscopy, supporting a CDW scenario as well.[33] In addition, although the CDW transitions are difficult to detect directly in most of the titanium oxypnictides, recent x-ray diffraction and Raman scattering measurements were able to reveal the structural modulation associated with the CDW transition in $Na_2Ti_2As_2O$.[60,61] Further experiments on high-quality single crystals are desirable for providing direct evidences of the possible CDW order in $Ba_2Ti_2Fe_2As_4O$.

# IV. Conclusions

Temperature-dependent inelastic neutron scattering (INS) measurements were performed on the self-doping superconducting material $Ba_2Ti_2Fe_2As_4O$. No magnetic signature could be evidenced in the attainable *(Q, E)* range of the experimental neutron energy loss setting. Phonon spectra were extracted from the INS measurements, and ab initio lattice dynamical calculations were performed to accompany the measured spectra. Our determined phonon spectrum and Raman frequencies from non-magnetic calculations agree well with our INS data and available Raman data[33]. Other calculations including spin-polarization and/or electron-correlation effects lead to phonon spectra deviating from the INS observations. This could be a signature of a destabilization of the crystal and electronic structures, in this case. Therefore, the non-magnetic framework describes the best the lattice dynamical behavior of $Ba_2Ti_2Fe_2As_4O$, which could be a step further towards establishing a CDW character of this material, especially that no experimental evidence is available on any magnetic signature stemming from the Fe or Ti sites.

[*]*zbiri@ill.fr*
[#]*w.jin@fz-juelich.de*